# X-Ray Emission from a prominent dust lane lenticular galaxy NGC 5866

N.D.Vagshette[1], S.S.Sonkamble[2], S.K.Pandey[3], M.K. Patil[2*]

[1]*IUCAA*,
Post bag 4, Ganeshkhind, Pune - 411 007 India
[2]*School of Physical Sciences*,
S.R.T.M. University, Nanded - 431 606, India
[3]*School of Studies in Physics*,
Pt. R. S. University, Raipur - 491 002, India
e-mail: patil@iucaa.ernet.in

We report the multiband imagery with an emphasis on the X-ray emission properties of a prominent dust lane lenticular galaxy NGC 5866. X-ray emission from this galaxy is due to a diffuse component and a substantial contribution from the population of discrete X-ray binary sources. A total of 22 discrete sources have been detected within the optical D25 extent of the galaxy, few of which exhibit spatial association with the globular clusters hosted by this system. Composite spectrum of the diffuse emission from this galaxy was well constrained by a thermal plasma model plus a power law component to represent the emission from unresolved sources, while that of the discrete sources was well fitted by an absorbed power law component of photon index $1.82 \pm 0.14$. X-ray color-color plot for the resolved source was used to classify the detected sources. The cumulative X-ray luminosity function of the XRBs is well represented by a power law function of index of $\Gamma \sim 0.82 \pm 0.12$.

Optical imagery of NGC 5866 revealed a prominent dust lane along the optical major axis of the host with dust extinction properties similar to those of the canonical grains in the Milky Way. The dust grains responsible for the extinction of starlight in NGC 5866 are relatively smaller in size when compared with the canonical grains in the Milky Way and high energetic charged particles seems to be responsible for the modulation of the dust grain size. Spatial correspondence is evident between the dust and other phases of ISM.

## 1 Introduction

Early-type galaxies, since their first time detection by the Einstein X-ray Observatory, have been regarded as the luminous X-ray systems with luminosities in the range from $10^{39} - 10^{42}$ erg/s (Forman et al. 1985). The bulk of the X-ray luminosity of this class of object was attributed to the hot ($\sim 10^7$ K) interstellar gas originating from the supernovae and winds of young massive stars hosted by it. However, a large dispersion was apparent in the X-ray luminosities of the early-type galaxies of a given range of optical luminosities. Further, spectral properties of the X-ray faint early-type galaxies were significantly different than those of their X-ray bright counterparts. Emission from the X-ray bright galaxies was found to be dominated by the thermal emission from hot interstellar medium (ISM) at kT $\sim 0.8$ keV, while that from the X-ray faint galaxies exhibited an additional hard ($\sim 5$-$10$ keV) component (Matsumoto et al. 1997). This additional, relatively harder component in the X-ray faint galaxies has characteristics similar to the sources seen in the bulge of the Milky Way and M31 and roughly follows the optical luminosity of their host (Irwin & Sarazin 1998). The potential source of emission for this additional component in the early-type galaxies was later identified as the population of discrete, bright X-ray point like sources, commonly known as low-mass X-ray binaries (LMXBs; Fabbiano 1989). Though, contribution from the latter component is significantly important, particularly in the case of X-ray faint galaxies, they have not been well identified before Chandra X-ray mission mainly due to the limitations in the spatial resolution of earlier generation X-ray observatories.

The unprecedented sub-arcsecond angular resolution of Chandra for the first time has enabled us to distinguish the emission from the discrete LMXBs relative to the surrounding diffuse emission. The LMXBs are comprised of a compact accreting primary (a neutron star or a black hole) and a low-mass main-sequence or red giant secondary that is losing material to the primary as a result of Roche lobe overflow. The compact primary object accretes matter from the evolved companion, which eventually gets heated and hence results in a copious amount of X-ray emission. Based on the mass of the accreting source, X-ray binary sources are classified as high-mass X-ray binaries (HMXBs) and low-mass X-ray binaries (LMXBs). Typical luminosities of the low mass X-ray binaries lie in the range of $\sim 10^{35} - 10^{39}$ erg s$^{-1}$. The excellent spatial resolution capability of *Chandra telescope* has been utilized extensively in studying the X-ray source population in distant galaxies since its launch (see Fabbiano & White 2006 for a review). Examples of the few systems in which this population has been investigated in great details include the Antennae galaxy (Fabbiano et al. 2001), M101 (Kuntz 2003), M81 (Pooley 2005), NGC4697 (Sarazin et al. 2001).

As hot gas distribution within a galaxy traces gravitational potential of the baryonic matter, therefore, X-ray emission maps derived for external galaxies provides a powerful tool in investigating structure of the galaxy. Further, morphology and extent of the hot gas in early-type galaxies provide important input on the nature of the hot gas, metal enrichment history of the ISM and hence formation scenario of the target galaxy. The diffuse gas detected in a galaxy through the X-ray emission is believed to originate from the supernovae and winds of young massive stars. Since supernovae generate metals, therefore, transport of the metal enriched hot gas is the key to understand metallicity evolution of host galaxy (e.g., Edmunds 1990). However, study of the purely diffuse component of the X-ray emitting gas was not possible before *Chandra* due to the lack of spatial resolution of the observing instruments. The superb angular resolution of *Chandra X-ray observatory* has not only allowed us to investigate contribution of discrete sources from diffuse emission component but has also allowed us to carry out precise morphological study of diffuse emission alone, thanks to the technological advancement. This has also enabled us to directly compare the morphologies of different phases of ISM, which otherwise was seriously impended due to the lack of higher resolution X-ray instruments.

In this paper we present the X-ray emission properties of a dust lane, edge on lenticular galaxy (S0$_3$) NGC 5866. The peculiar property of NGC 5866 is that it hosts a remarkable dust lane oriented along its optical major axis at a redshift of $v_r = 824$ km/s. NGC 5866 has two companions, NGC 5907 (Sc; $v_r = 779$ km/s) and NGC 5879 (Sb; $v_r = 929$ km/s). The study of NGC 5866 in X-ray bands in this paper is supplemented by the ancillary data at other wavelengths e.g. optical broad band and narrow band acquired using IGO, Pune and UV and near-IR data from the archives of space observatories. Main objective of the present paper is to understand the nature and properties of X-ray emission, investigate its association with other phases of the ISM and to explore the population of the resolved sources in NGC 5866. This galaxy has been reported to harbor a sizable number of globular clusters (GCs), therefore, it is interesting to investigate the population of discrete sources and also to check their association with the globular clusters. Excess emission at 6.7 and 15 μm has been detected in the case of NGC 5866 meaning that NGC 5866 is passing through the star burst phase (Xilouris et al. 2004). As a result, NGC 5866 acts as an ideal target to study the association of GCs with the LMXBs as well as the dust lane.

This paper is organized as: Section 2 describes observations and data preparation steps of the X-ray as well as optical broadband images on NGC 5866. Section 3 discusses the results from the imaging analysis of the X-ray and optical photons with a view to investigate dust extinction proper- ties in the target galaxy. Discussion on the results is provided in the Section 5, while we summarize our study in Section 6.

**2 Observations and Data Preparation**

*2.1 X-ray Data*

NGC 5866 was observed with the back-illuminated Chandra ACIS-S3 detector on 14 November 2002 for a total exposure of 36.7 ks (ObsID 2879). We acquired the level-1 event file on this target from the archive of Chandra observatory. The data were reprocessed in a uniform manner using the standard tasks available within the Chandra Interactive Analysis of Observations (CIAO) v.4.2.0 and the corresponding calibration files CALBD v.4.3.0 provided by the Chandra X-ray Centre (CXC). Periods of high background were verified by plotting its 0.3 - 10 keV light curve extracted from the chip and were filtered out using the 3σ clipping. After removal of the background flares, the clean data had a net exposure of 30.77 ks. Background subtracted, exposure corrected, adaptively smoothed 0.3 - 10 keV Chandra emission map of NGC 5866 is shown in Figure 1.

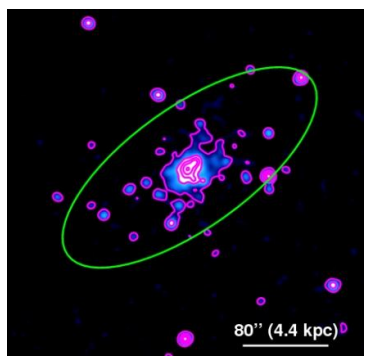

Fig. 1: 0.3-10 keV background subtracted, exposure corrected, adaptively smoothed Chandra X-ray image delineating hot gas distribution within NGC 5866. Positions of the resolved point sources within chip S3 are also shown.

*2.2 Optical Data*

Deep CCD images on the target galaxy NGC 5866 were obtained using the 2.0 m Optical Telescope of IUCAA Girawali Observatory (IGO), Pune, India during March 2010 in the Bessels B, V, R & I broad band filters and the narrow band filter centered on the Hα emission. The IUCAA Faint Object Spectrograph Camera (IFOSC) equipped with EEV 2k×2k thinned, back-illuminated CCD with 13.5 μm pixels was used for this purpose. The CCD used for imaging provides an effective $10.'5 \times 10.'5$ field of view on the sky corresponding to a plate scale of $0.''3$ $pixel^{-1}$. The gain and read out noise of the CCD camera are 1.5 $e^-$/ADU and $4e^-$, respectively. Multiple exposures in each of the broad band as well as the narrow band Hα filter were taken to achieve good signal-to-noise (S/N) ratio. For the calibration of science frames we have also acquired calibration frames like, bias, twilight sky frames during this observing run.

The optical observations acquired during this run were analyzed using the Image Reduction and Analysis Facility (IRAF; *http://stsci.harvard.edu/IRAF*) software and following the standard preprocessing steps like, bias subtraction, flat fielding, etc. (see Patil et al. 2007 for details). Multiple frames taken in each pass band were co-added after geometrically aligning these images to accuracy better than one tenth of a pixel by measuring centroids of several common stars in the galaxy frames. Sky background in the individual band pass was estimated using box method and was then subtracted from the respective frame. These cleaned, sky subtracted images were used for delineating morphology of the interstellar dust as well as for deriving dust extinction properties in NGC 5866.

**3 Results**

*3.1 Properties of X-ray emitting gas*

As stated earlier morphology, spatial extent and quantum of the hot gas in a galaxy provides an important input on the nature of the hot gas, metal enrichment history of the ISM and hence formation scenario of the galaxy. Therefore, we derive a pure diffuse X-ray emission map due to the diffuse gas alone from the adaptively smoothed, exposure corrected, point source removed X-ray image of NGC 5866. The point sources within S3 chip were detected using the task *wevdetect* within CIAO on the full-band *Chandra* image. Those detected areas were removed and the 'holes' due to the removal of point sources were 'filled-in' by extrapolating the local backgrounds using the Poisson's method. The resultant adaptively smoothed X-ray emission map of NGC 5866 in the soft energy band (0.3-2 keV) is shown in Figure 2. A casual inspection of this figure reveals that the X-ray emission from this system is distributed throughout the galaxy roughly following the stellar distribution within it. In the same figure we also show the X-ray emission maps in the medium energy band (1-2 keV) and the hard band (2-10 keV) derived in a similar way from the cleaned Chandra image. Comparison of these images reveal that the hard component is originating from the nuclear region only, while the soft component follows the morphology of optical light distribution meaning that it has mostly originated from stellar sources. Thus, the diffuse X-ray emission from NGC 5866 particularly in the soft band has the ellipticity and position angle same as that of the optical emission whereas that at harder energy band appears to be rounder and confined to the central portion only.

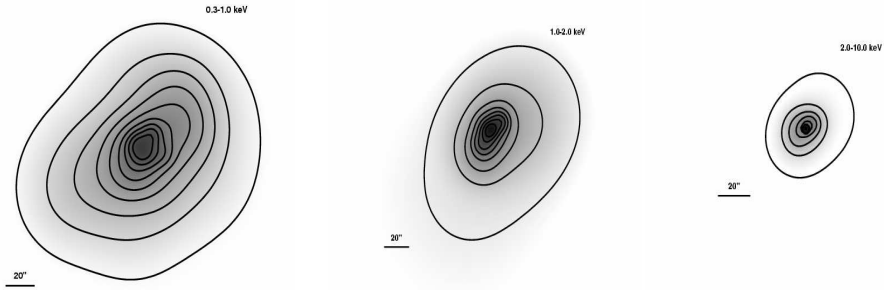

Fig. 2: Adaptively smoothed, exposure corrected, point sources removed diffuse emission maps of NGC 5866 in the soft (0.3-1 keV, *left panel*), med (1-2 keV, *middle panel*), and the hard (2-10 keV, *right panel*) energy bands. Iso-intensity contours in different energy bands have been overlaid on these images.

To investigate the 2D distribution of hot gas within NGC 5866, we compute azimuthally averaged surface brightness profile of the diffuse X-ray emission in 0.3-3 keV as a function of radial distance. For this we extract X-ray flux values from narrow circular concentric annuli centered on the X-ray peak of NGC 5866. During this extraction regions surrounding the resolved point sources have been excluded. The data points were then fitted with the standard two-component 1-d *beta* model of varying power law using the task *beta1d* available within the CIAO *Sherpa*, which is empirically given as

$$\sum(r) = \sum_0 \left[1 + \left(\frac{r}{r_c}\right)^2\right]^{-3\beta+0.5}$$

where $\Sigma_0$, $r_c$ and $\beta$ respectively represent the central X-ray surface brightness, core radius and slope of the surface brightness profile. Figure 3 displays surface brightness profile of the X-ray emitting hot gas distribution as a function of projected distance r. In this figure we also plot the best-fit standard 1D β model (continious line), which resulted in the best fit parameters equal to β=0.62 and $r_c$=9.″85 (640 pc).

To investigate the global properties of hot gas distribution within NGC 5866, we extracted a point source removed, background subtracted 0.3-7 keV spectrum of the diffuse emission from within the optical D25 ellipse of NGC 5866. We excluded central 2 arcsec region to avoid contamination due to the nuclear source. The background spectrum was extracted from the exposure corrected blank sky frame downloaded from the CXC. To account for the variations in the normalization of the instrumental background, we normalized the system supplied background files using the 10 - 12 keV count rates in the science frames. Realizing that this spectrum is representing emission from the diffuse hot gas alone, we treated this spectrum by fitting it with an optically thin thermal plasma emission model APEC in XSPEC modified by the Galactic absorption component fixed at $1.38 \times 10^{20}$ cm$^{-2}$ and following the standard $\chi^2$ statistics. Temperature, metal abundance and normalization were allowed to vary during the fit. However, the fit revealed residuals particularly in the high energy range. This may be due to the contribution from the unresolved point sources, which exhibit hard spectrum. Therefore, to constrain this spectrum in a proper way we added a power law component to it, which resulted in a relatively better fit with $\chi^2$ value close to 1.2. Our best-fit model resulted into the hot gas temperature kT~0.19, the power-law index of $\Gamma \sim 1.82$ (Table 1), and an abundance of Ne = $0.16^{+0.57}_{-0.45}$ and Fe = $0.70^{+0.76}_{-0.38}$ in solar units. The additional power law component imply that the resolved point source removed X-ray emission from NGC 5866 is contaminated by a population of unresolved LMXBs and fainter compact objects that remained undetected earlier.

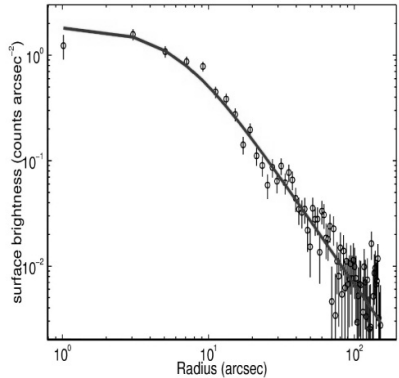

Fig 3: Azimuthally averaged, background subtracted 0.3-3 keV radial surface brightness profile of the diffuse X-ray emission from NGC 5866 along with the best-fitted β model (continuous line)

Table 1: Spectral Fit for NGC 5866

| Component | kT/Γ (keV/—) | $N_e$ (solar) | $F_e$ (solar) | $L_{res.}$ (ergs s$^{-1}$) | $L_{unres.}$ (ergs s$^{-1}$) | $L_{diff}$ (ergs s$^{-1}$) | $L_{total}$ (ergs s$^{-1}$) | chisq/dof |
|---|---|---|---|---|---|---|---|---|
| res. srcs. | —/$1.82^{+0.14}_{-0.14}$ | — | — | $1.02 \times 10^{39}$ | — | — | — | 1.02 |
| diff.+unres. | $0.18^{+0.05}_{-0.03}$/(1.82) | $0.18^{+2.1}_{-0.13}$ | $0.41^{+0.63}_{-0.28}$ | — | $4.98 \times 10^{38}$ | $1.20 \times 10^{39}$ | $1.69 \times 10^{39}$ | 0.93 |
| total | $0.19^{+0.16}_{-0.039}$/(1.82) | $0.16^{+0.57}_{-0.45}$ | $0.70^{+0.76}_{-0.38}$ | — | — | — | $3.15 \times 10^{39}$ | 1.20 |

We have also extracted a spectrum of the total X-ray emission including that originating from the resolved sources from within one effective radius of NGC 5866. The composite spectrum was then treated with an absorbed thermal and power-law component (APEC+power-law), which resulted in the best fit temperature value of kT $\sim 0.19^{+0.16}_{-0.039}$ keV and the power law index of 1.82 (Table 1). Thus, the resultant fit for this composite spectrum is consistent with the best-fit spectra due to resolved and unresolved emission independently.

*3.2 X-ray binary sources*

X-ray point sources recorded on chip S3 were detected using the CIAO tool *wevdetect* in the energy band 0.3 - 8 keV. The detection algorithm searched for the sources over pixel scales 1, 2, 4, 8, and 16 pixels with the detection threshold set to $\sim 10^{-7}$, equivalent to only one spurious detection per image. This tool has a capability to separate closely spaced point sources. The point sources identified by *wevdetect* were verified by visual inspection of the image so as to avoid the obvious errors made during the detection. A total of 72 X-ray binary sources (XRBs) were detected on the S3 chip, of which 30 are lying within the B-band 25$^{th}$ magnitude isophote $D_{25}$ of NGC 5866. However, out of the 30 sources only 22 exhibit a minimum signal-to-noise (S/N) ratio of 3. Therefore, we used only these 22 point sources within optical $D_{25}$ for further exploration and studying their population in NGC 5866.

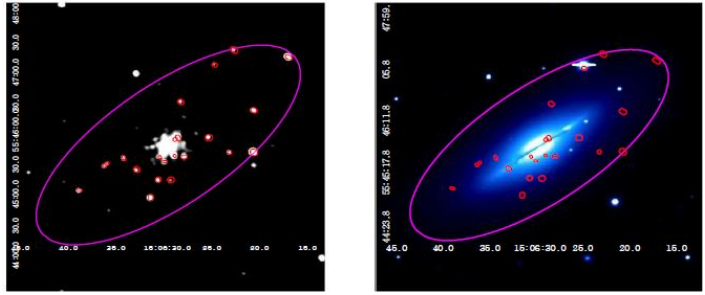

Fig 4: (left panel) Resolved point sources as well as diffuse X-ray emission from NGC 5866. This image has been smoothed with a Gaussian of 3σ. (right panel) Regions of the resolved point sources have been overlaid on the DSS *B-band* image to confirm that these sources are not coinciding with the foreground stars. The purple ellipses in both the figures delineate the optical $D_{25}$ extent of NGC 5866.

Figure 4 (left panel) shows both resolved point sources as well as the diffuse X-ray emission from NGC 5866. For the best representation, we have smoothed this image using a minimum of 3 sigma signal-to-noise ratio (S/N) per smoothing bin. This figure clearly shows that majority of the X-ray emission in NGC 5866 is originated from the point sources and the contribution due to the hot diffuse component is very little. To examine the spatial distribution of the detected point sources and to confirm that they are not due to the foreground stars we overlay regions of the resolved sources on to the Digital Sky Survey (DSS) *B-band* optical image of NGC 5866, shown by red circles in Figure 4 (right panel). This figure confirms that none of the detected sources coincide with the foreground stars.

Barring few exceptions, X-ray counts from majority of the resolved sources are not enough to fit their spectra individually. As a result it is very difficult to investigate spectral properties of the individual source and hence to classify them on the basis of their spectral properties. However, X-ray colors or hardness ratios of these sources will enable us to understand the crude spectral characteristics of the individual source (Sivakoff et al. 2008). Following Sarazin et al. (2000), we extract source counts in three different energy bands namely, soft (S; 0.3-1.0 keV), medium (M; 1.0-2.0 keV) and the hard (H; 2.0-5.0 keV). Then we estimate hardness ratios of individual source using the definitions H21 = (M-S) / (M+S) and H31 = (H-S) / (H+S). The hardness ratios and associated 1σ errors of all the 22 point sources within optical $D_{25}$ region of NGC 5866 are given in columns 8 and 9 of Table 2 and are also plotted in Figure 5 (left panel). The overall nature of the source colors in NGC 5866 is similar to those seen in other dusty early-type galaxies as reported by Vagshette et al. (2013) and reveals that majority of the sources within NGC 5866 have negative hardness ratios indicating that they mostly emit at lower energies. There is only one source with hardness ratio (H21, H31) ~ (1.0, 1.0) and probably represent the central source of NGC 5866. Similarly, there are two more source that have X-ray colors (+0.5, +0.5) and exhibit hard emission in the range 2-7 keV relative to the soft band. This system also hosts a source whose hardness ratio (H21, H31) ~ (-0.7, -1.0) and is softest among the detected sources. We do not find any source in NGC 5866 which has harder X-ray color i.e., source with hardness ratio (H21, H31) > (1.0, 1.0). Majority of the remainders are like the normal neutron star accreting LMXBs. Table 2 lists properties of all the 22 sources detected within the optical $D_{25}$ region of NGC 5866 and are listed according to their position coordinates, projected distance from the optical center of the galaxy, count rate, significance of their detection, unabsorbed X-ray luminosity and X-ray soft and hard colors of the detected sources, respectively. In the same figure (right panel) we also plot the modified X-ray colors of the resolved sources, derived after normalizing the X-ray colors by the X-ray counts in the total band (0.3 - 10 keV). A casual look at this figure clearly helps us in classifying the resolved sources in different types depending on their emission characteristics (Vagshette et al. 2012). Thus, there is a very heavily absorbed hard source among the detected sources and probably it is the nuclear source. There is one very soft source mostly emitting in lower energy band (0.3 - 2 keV). Rest of the sources are the normal neutron star accreting LMXBs.

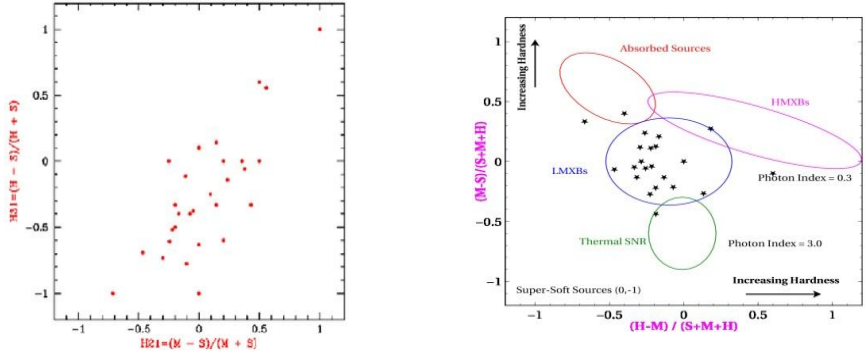

Fig. 5: (left panel :) X-ray color-color plot of all the resolved sources within optical $D_{25}$ extent of NGC 5866. The X-ray colors have been defined as H21=(M-S)/(M+S) and H31=(H-S)/(H+S), where S, M and H are the X-ray counts in the soft (0.3-1.0 keV), med (1.0-2.0 keV) and hard (2.0-5.0 keV) energy bands, respectively. (right panel:) Modified X-ray color diagram of the resolved stars used to classify them, overlaid on it are the regions of different classes of the XRBs.

Table 2: Properties of the X-ray binary sources resolved within optical $D_{25}$ of NGC 5866

| Source No. | R.A. h:m:s | Dec °:':" | D (kpc) | Count Rate $10^{-4}$ cts s$^{-1}$ | SNR ---- | $L_X$ $10^{37}$ erg s$^{-1}$ | H21 ---- | H31 ---- |
|---|---|---|---|---|---|---|---|---|
| (1) | (2) | (3) | (4) | (5) | (6) | (7) | (8) | (9) |
| 1 | 15:06:31.270 | +55:44:58.10 | 2.54 | 13.1±2.10 | 06.24 | 8.19±1.310 | -0.14 | -0.67 |
| 2 | 15:06:38.728 | +55:45:05.48 | 4.32 | 7.39±1.57 | 04.69 | 4.62±0.985 | -0.30 | -0.56 |
| 3 | 15:06:32.726 | +55:45:25.50 | 1.70 | 5.04±1.30 | 03.87 | 3.15±0.814 | -0.17 | -0.25 |
| 4 | 15:06:35.792 | +55:45:31.71 | 2.69 | 3.02±1.01 | 03.00 | 1.89±0.630 | 0.14 | -0.33 |
| 5 | 15:06:30.312 | +55:45:37.64 | 0.59 | 3.36±1.06 | 03.16 | 2.10±0.664 | -0.33 | 0.75 |
| 6 | 15:06:27.756 | +55:45:37.87 | 0.89 | 7.05±1.54 | 04.58 | 4.41±0.963 | -0.05 | -0.64 |
| 7 | 15:06:28.757 | +55:45:38.83 | 0.55 | 14.1±2.18 | 06.48 | 8.82±1.360 | 0.31 | -0.35 |
| 8 | 15:06:23.017 | +55:45:42.11 | 2.72 | 5.37±1.34 | 04.00 | 3.36±0.840 | -0.47 | -0.60 |
| 9 | 15:06:20.487 | +55:45:42.28 | 3.76 | 43.0±3.80 | 11.31 | 26.9±2.380 | -0.25 | -0.41 |
| 10 | 15:06:28.721 | +55:45:55.01 | 0.5 | 14.8±2.23 | 06.63 | 9.24±1.390 | 0.00 | 0.00 |
| 11 | 15:06:25.183 | +55:45:56.63 | 1.86 | 5.71±1.38 | 04.12 | 3.57±0.866 | 0.14 | -0.45 |
| 12 | 15:06:20.465 | +55:46:22.69 | 4.13 | 7.72±1.61 | 04.80 | 4.83±1.010 | 0.29 | -0.22 |
| 13 | 15:06:28.105 | +55:46:31.57 | 2.23 | 4.70±1.26 | 03.74 | 2.94±0.786 | 0.00 | -0.50 |
| 14 | 15:06:16.829 | +55:47:15.23 | 6.79 | 25.2±2.91 | 08.66 | 15.8±1.820 | -0.05 | -0.38 |
| 15 | 15:06:30.475 | +55:45:15.58 | 1.62 | 5.04±1.30 | 03.87 | 3.15±0.814 | -0.40 | 0.25 |
| 16 | 15:06:29.155 | +55:45:15.29 | 1.6 | 3.02±1.01 | 03.00 | 1.89±0.630 | 0.33 | -1.00 |
| 17 | 15:06:36.105 | +55:45:29.27 | 2.85 | 4.03±1.16 | 03.46 | 2.52±0.728 | 0.60 | 0.20 |
| 18 | 15:06:29.808 | +55:45:33.17 | 0.72 | 4.70±1.26 | 03.74 | 2.94±0.786 | -0.27 | -0.14 |
| 19 | 15:06:34.072 | +55:45:36.66 | 1.94 | 5.37±1.34 | 04.00 | 3.36±0.840 | 0.17 | -0.27 |
| 20 | 15:06:28.403 | +55:45:56.45 | 0.64 | 10.7±1.90 | 05.66 | 6.72±1.190 | -0.07 | -0.47 |
| 21 | 15:06:24.573 | +55:47:07.79 | 4.43 | 3.69±1.11 | 03.32 | 2.31±0.697 | 0.50 | -0.50 |
| 22 | 15:06:22.534 | +55:47:22.28 | 5.47 | 5.04±1.30 | 03.87 | 3.15±0.814 | -0.07 | -1.00 |

Notes: Columns (2) & (3) give position of the source; (4) gives source distance from center of galaxy; count rates converted to luminosity assuming conversion factor $6.258 \times 10^{40}$ erg count$^{-1}$

As X-ray counts from individual source were not enough to perform their spectral analysis, therefore we extracted a 0.3-8 keV composite spectrum of all the resolved sources within the optical $D_{25}$, except the central source. Background spectra were extracted from the local region surrounding individual source. This composite spectrum was then treated using a signal-component power-law model holding the absorption fixed at Galactic value ($1.38 \times 10^{20}$ cm$^{-2}$) by exporting it to XSPEC (V12.6.0). This resulted in the best-fit power law index of $\Gamma \sim 1.82$ and the reduced $\chi 2$ little higher than one. This fit revealed residuals in the lower energy range. Therefore, we tried to constrain this composite spectrum by including a thermal component to the power-law model and the fit showed a significant improvement with the reduced $\chi 2$ reaching to one. This additional soft component probably represent the super-soft sources from our detection that has very soft X-ray colors (Vagshette et al. 2013).

*3.3 Dust morphology and extinction*

This galaxy has long been known to host a prominent dust lane oriented almost parallel to its optical major axis. However, to examine its extent and compare its morphology and properties with the X-ray emission from NGC 5866, we analyzed cleaned broadband optical images of NGC 5866. To delineate the accurate morphology and extent of the dust lane we derived its color index images like, (B-V), (B-R), etc. by comparing light distribution in two different passbands after convolving them with the Gaussian function so as to match their *psfs*. One of such color index maps (B-V) is shown in Figure 6(a), which clearly reveals a prominent dust lane aligned along the optical major axis of NGC 5866.

To quantify the wavelength dependent nature of dust extinction in NGC 5866 we generate its dust free models in each pass-band by fitting ellipses to the iso-intensity contours of the observed images (see Patil et al. 2007 for details). For this purpose we have used the task *ELLIPSE* available within STSDAS. The dust occupied regions as well as regions occupied by the foreground objects were masked and flagged off during the fit. Because of the prominent dust lane running along the semi major axis of NGC 5866, fitting ellipses to the isophotes was not an easy task. We, therefore, fitted ellipses to the isophotes in longer wavelength image first and then repeated for the shorter wavelength. This ellipse fitting enabled us to derive dust free models of NGC 5866 in each pass band, which were then used to derive the extinction maps (Patil et al. 2007). Figure 6(b) shows one of such extinction maps derived in V band for NGC 5866, which clearly exhibits a prominent dust lane running along the optical major axis of the target galaxy. Similar extinction maps were also derived for the images in near-IR (J, H & Ks) bands, taken from the archive of 2MASS observatory, before which they were scaled to match with the images in optical pass bands for their different plate scales.

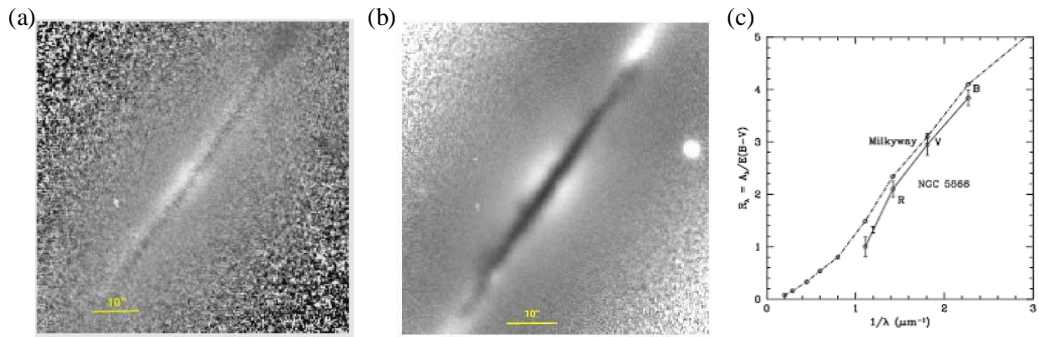

Fig. 6: (a) (B-V) color index map of NGC 5866 derived by comparing light distribution in properly convolved two different passband images. (b) V band dust extinction map derived after comparing light distribution in the original image with that in the dust free model derived from ellipse fitting technique. Dark shades in both the images delineate the dust occupied region. (c) Extinction curve for NGC 5866 (continuous line) plotted as a function of inverse of the wavelength. For comparison standard Galactic extinction curve is also plotted (dash-dotted line).

We then quantify the total extinction $A_\lambda$ in each passband ($\lambda$=B, V, R & I) and use them to plot the extinction curve (Patil et al. 2007). The extinction curve for NGC 5866 is shown in Figure 6 (c, continuous line) and is found to run parallel with that for the Milky Way (dash-dotted line), implying that the extinction properties of the dust grains in the extragalactic environment are identical to those of the canonical grains in the Milky Way. The $R_\lambda$ value (ratio of the total extinction in V band to the selective extinction in B and V bands) varies linearly with the inverse of wavelength and is consistent with the results derived by Goudfroij et al. (1994), Patil et al. (2007), Deshmukh et al. (2012), Kulkarni et al. (2014). The relative grain size estimated by shifting the observed extinction curve for NGC 5866 along $\lambda^{-1}$ axis until it best matches with the Galactic extinction curve is found to be equal to $<a>/<a_{Gal}> = 0.90$.

Assuming the spherical, two-component model of silicate and graphite grains with an adequate mixture of sizes (Mathis, Rumpl & Nordsieck 1977) and measuring the total extinction in V band image, we compute the dust content of NGC 5866 to be equal to $2.5 \times 10^4$. Note that this method accounts only for the dust responsible for the extinction of the visual light in NGC 5866; any component that is diffusely distributed within the galaxy is not sensitive to the extinction of the stellar light and hence is not accounted for in this method. As a result this visual extinction method always provide a lower limit to the estimate of total dust content of a galaxy (Patil et al. 2007). The true content of dust within NGC 5866 can be quantified by taking the flux densities measured at longer wavelengths i.e., flux densities from IRAS or ISO satellites. We estimate the dust temperature as well as dust mass using the IRAS flux densities at 60 μm and 100 μm and are found to be equal to 30 K and $7.6 \times 10^5$ M$_\odot$, respectively (Young et al. 1989). This means dust mass estimate from the IRAS flux density is higher by an order of magnitude relative to that from the dust screening in optical passbands.

*3.4 Multiband Imagery*

Optical images acquired for NGC 5866 revealed a prominent dust lane along the optical major axis. With an objective to examine the spatial correspondence between the X-ray emitting hot gas with that of the interstellar dust, ionized gas and stellar distribution mapped through the Ks band image, we produce multi-wavelength images of NGC 5866. Figure 7 shows multiband imagery of the dust lane lenticular galaxy NGC 5866 mapped in (a) soft 0.3-1 keV X-ray band, (b) optical B-band image acquired using the 2.0m IGO telescope, (c) the GALEX Near-UV band and (d) stellar emission mapped in Ks Near-IR band using the 2MASS telescope. For better visualization of the hot gas within NGC 5866 the 0.3 - 1 keV X-ray image has been smoothed using an adaptive smoothing tool *csmooth* with a minimum 3 sigma S/N ratio. This imagery clearly reveals the morphological similarity among all the four different components. To map distribution of the ionized gas within NGC 5866 we derived its Hα emission map using the narrow images acquired using the IGO 2.0m telescope and is shown in Figure 8(a). For this we have scaled the narrow band image with intensity distribution in R-band image by fitting ellipses to the outer parts and then the properly scaled R-band image was subtracted from the narrow band image centered on the Hα emission. This figure reveals that the Hα emission is mostly confined to the dust occupied region (lane) relative to the extended X-ray emission. In Figure 8(b) we plot radial profiles of the surface brightness distributions in the 0.3 -1 keV X-ray band (shown by the red dashed line), optical B-band (blue continuous line), GALEX Near-UV band (green dotted line) and the 2MASS Near-IR Ks band (long-dashed magenta line). For this we have extracted counts from the concentric elliptical annuli centered on the peak of the X-ray emission from NGC 5866. This figure depicts similarity in the emission properties of different components. Dips in the X-ray and optical profiles represent the loss of photons due to the absorption by dust grains and is maximum near the center.

As mid-IR images centered on 3.6 μm, 8 μm and 24 μm acquired using the Infrared Array Camera (IRAC) of the Spitzer Space Telescope (SST) can potentially trace the stellar mass distribution of external galaxies, therefore we have made use of these data sets for the present study and are shown in Figure 9. A comparison of the morphology of the dust emission mapped through these bands exhibits a spatial correlation with that of the X-ray as well as ionized gas emission maps.

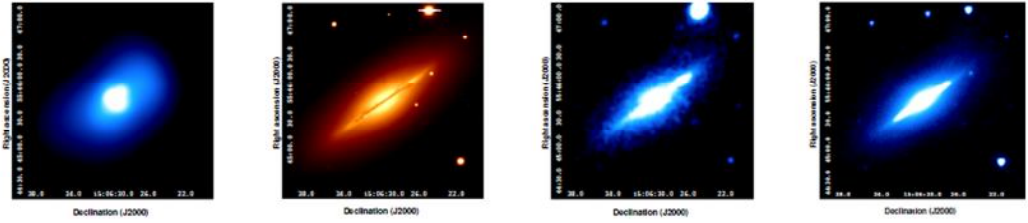

Fig. 7: Multiband imagery of the dust lane early-type galaxy NGC 5866 in (a) 0.3-1 keV soft X-ray band, (b) Optical B-band, (c) GALEX Near-UV band and (d) 2MASS Ks band.

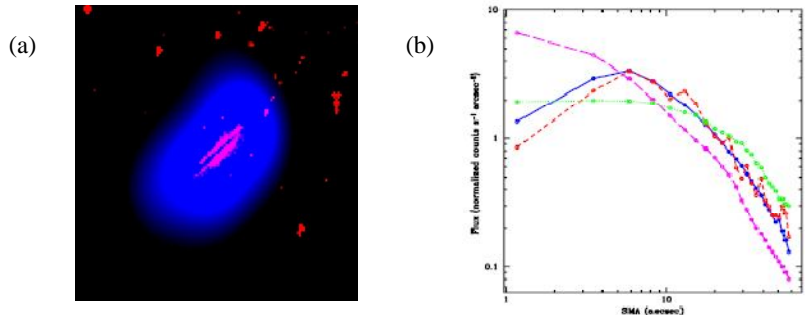

Fig. 8: (a) Chandra 0.3 - 1 keV diffuse X-ray emission from NGC 5866 (shown blue color) compared with the ionized gas morphology (shown in magenta color). Notice ionized gas is confined to the dust disk only.
(b) Radial profiles of the surface brightness distributions in the 0.3 - 1 keV X-ray band (shown by the red dashed line), optical B-band (blue continuous line), GALEX Near-UV band (green dotted line) and the 2MASS Near-IR Ks band (long-dashed magenta line)

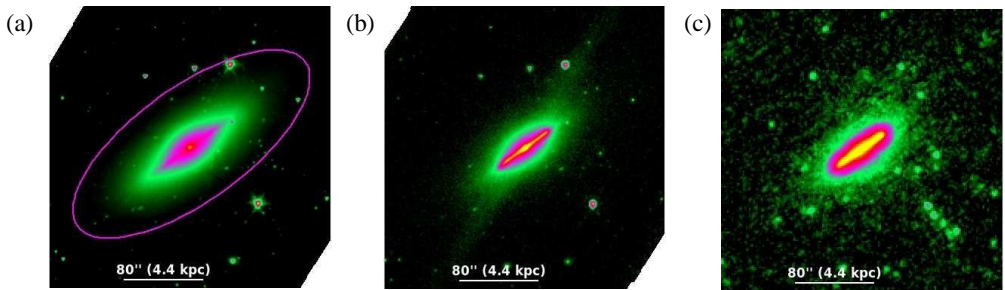

Fig. 9: Dust emission maps for NGC 5866 in the SPITZER IRAC passbands (a) at 3.6 µm (b) 8 µm (c) 24 µm

*3.5 XRBs vs. Globular clusters*

Important feature of the LMXBs as regard to their formation is provided by their possible association with the globular clusters (GCs). It has been proposed that the XRBs in early-type galaxies are mostly formed in the globular clusters, the regions where the stellar densities are sufficiently large for the formation of XRBs through the multi-body interactions (White et al. 2002). Systematic studies of nearby, massive elliptical galaxies have revealed that about 30-60% of the luminous XRB sources are associated with the globular clusters hosted by these galaxies. Further, X-ray studies of these globular clusters have revealed that about 4% of the GCs host LMXB sources (Sarazin et al. 2001, Kraft et al. 2005). It has also been evidenced that normally brighter and redder GCs are the most favored place for the association of LMXBs. This means that there is an important relationship between the LMXBs and the GC population at least in the most massive early-type galaxies. To map the GCs in galaxies, the globular cluster specific frequency (GCSF) function is strongly preferred which depends on environment as well as morphological type of the host galaxy. Elliptical galaxies generally have larger GCSF relative to the lenticular galaxies. The GCSF for NGC 5866 has been reported to be 1.4 indicating that there are about 109 GCs associated with this system (Cantiello et al. 2010).

With an objective to examine association of the detected XRB sources with the globular clusters in NGC 5866, we compare positions of the LMXBs with those of the GCs and are shown in Figure 10(a). This comparison revealed that out of the 22 detected sources only 2 exhibit their association with the GCs. Off the remainder, 4 are within 1″ of the globular clusters while 7 are about 3″ away from the GCs. This study revealed that about 9 XRBs in NGC 5866 can be accounted for in the 109 GCs in 2 arcmin radius or 0.32 XRBs per GC. Thus, a significant fraction of LMXBs in this galaxy are found to be associated with the globular clusters, implying that majority of them have formed in the globular clusters via stellar dynamical interactions. It is believed that the LMXBs whose association with GCs has not been confirmed may also have formed in the GCs, but have kicked off after their formation due to the supernova explosions.

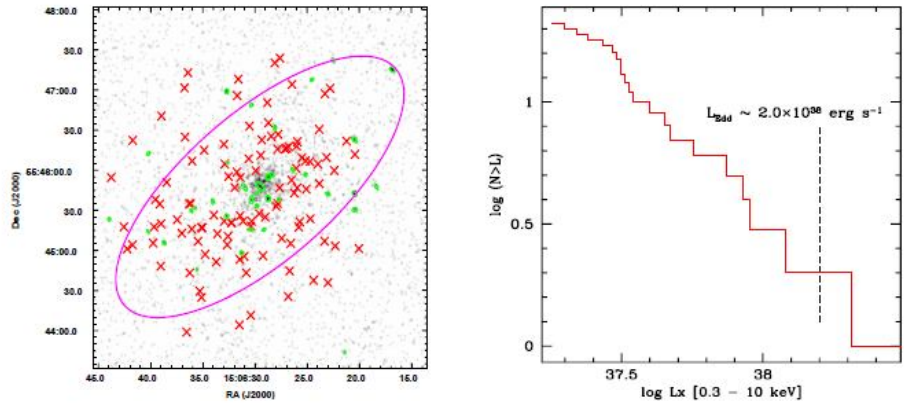

Fig. 10(a): Association of XRBs with the GCs within optical $D_{25}$. Red crosses indicate the positions of the GCs, while that of the XRBs are shown by green circles. (b) X-ray Luminosity Function (XLF) of all the detected 22 XRBs within $D_{25}$ of NGC 5866

Assuming that all the detected sources within optical $D_{25}$ are located at the distance of NGC 5866 (11.3 Mpc), we converted the source count rates into the unabsorbed 0.3 - 10 keV X-ray luminosities. For this we used the best-fit *Chandra* power law spectrum of the resolved point sources. This resulted in the conversion factor of $5.42 \times 10^{40}$ erg cts$^{-1}$. The resulting 0.3 - 10 keV X-ray luminosities of the resolved sources are found to range between $\sim 1.89 \times 10^{37}$ to $\sim 2.69 \times 10^{38}$ erg s$^{-1}$. To characterize different populations of the resolved XRBs within NGC 5866 we obtain the X-ray luminosity function (XLF) of all the sources using their estimated X-ray luminosities. Figure 10(b) reveals the uncorrected-cumulative luminosity function for discrete sources in NGC 5866. To build up the XLF, we restricted to the detection limit of S/N > 3 within the $D_{25}$ boundary of the galaxy. We tried to fit this cumulative luminosity function with a simple power law, which was acceptable with ∼90% confidence and the best fit power law index of $\Gamma = 0.82 \pm 0.12$. The Eddington limit $L_{X,Edd} \approx 2 \times 10^{38}$ erg s$^{-1}$ for a 1.4 $M_{\odot}$ neutron star is marked by the vertical dashed line in this figure. Only one source from our detection is found to have X-ray luminosity higher than the Eddington limit.

**4 Discussion**

Multiband optical imagery of NGC 5866 enabled us to investigate the dust morphology and dust extinction properties in the extragalactic environment. The dust extinction properties in NGC 5866 were found to be identical to those of the canonical grains in the Milky Way as inferred from the nature of the extinction curve. However, the $R_V$ value, which characterizes physical size of the dust grains takes a value little smaller than that for canonical grains. The dust grains in NGC 5866 appear to have been modulated due to the interactions with the relativistic charged particles (Trincheri & Goudfrooij 2002).

Using the measured values of dust extinctions in V band image we quantify total dust content of NGC 5866 to be equal to $1.0 \times 10^4$ $M_\odot$ and was found to be lower than $8.9 \times 10^5$ $M_\odot$, estimated using the IRAS flux densities at 60 and 100 μm. This gives the ratio of the two estimates, $M_{d,IRAS}/M_{d,optical}$ ~89. Discrepancy in the two estimates is due to the fact that the dust mass derived from optical extinction assumes foreground screening and hence is not sensitive to the component of the dust embedded within the ISM (Patil et al. 2007). IRAC observations of NGC 5866 at 8μm further enhanced this discrepancy with a total dust content of ~ $4.5 \times 10^6$ $M_\odot$ (Draine et al. 2007). Narrow band images centered on the Hα emission from this galaxy showed a spatial correspondence of the ionized gas with the interstellar dust, whose total content is found to be equal to ~ $4.7 \times 10^7$ $M_\odot$.

Origin of dust and ISM in early-type galaxies is a highly controversial issue. However, there is a consensus that it might have originated either internally from the mass-loss of the evolved stars and/or externally due to the accretion of matter during the tidal capture or merging of gas rich galaxies. Observational evidences of a large sample of dust lane early type galaxies have favored the external origin of the dust and ISM (Bertola et al. 1992, Goudfrooij et al. 1994, Patil et al. 2007). The dust grains once produced, are subjected to the exotic environment of constant bombardment by the high energetic charged particles. As a result the dust grains in the hostile environment must be undergoing size modulation due to the erosion and sputtering by supernova blast waves, grain-grain collisions, sputtering by thermal ions (warm and hot), heating by hot electrons, etc. Therefore, dust grains in such cases must be smaller than the canonical grains in the Milky Way. We used the 1D beta-model fit to the X-ray surface brightness distribution from NGC 5866 to investigate the physical properties of the hot gas i.e., electron density and hence to estimate the total hot gas content of NGC 5866 (Canizares et al. 1987). This resulted in to the central electron density of $n_e(0) = 3.6 \times 10^{-2}$ cm$^{-3}$ and the total mass of the hot gas within $D_{25}$ ellipse equal to ~ $3.0 \times 10^7$ $M_\odot$. These estimates of electron density were then used to derive the dust grain destruction time scale using:

$$\tau_d = a \left| \frac{da}{dt} \right|^{-1} \approx 2 \times 10^4 \left( \frac{cm^{-3}}{n_H} \right) \left( \frac{a}{0.01 \mu m} \right) yr$$

where $n_H$ is the proton density ($n_H = 0.83\ n_e$) in the plasma and $a$ is the grain size. Using this value of dust grain destruction time scale and the mass injection due to the evolved stars and supernovae, we estimate the total dust content of NGC 5866 to be equal to $1.17 \times 10^4$ $M_\odot$, much smaller than that estimated from visual extinction. This means NGC 5866 might have acquired the visible quantum of dust through some merger like episode.

**5 Summary**

We present the X-ray emission and the population of discrete sources in the prominent dust lane lenticular galaxy NGC 5866. This study is also supplemented by the multi-wavelength imagery of NGC 5866 with an emphasis to examine association of dust and X-ray emission from NGC 5866. Important results from this study are summarized below:
- 33.7 ks high resolution Chandra image on NGC 5866 enabled us to investigate the X-ray emission properties and also to delineate the population of X-ray binary sources in this prominent dust lane, star forming lenticular galaxy.

- Soft X-ray emission originating from the hot gas alone is found to coincide with the central part of the galaxy following morphology of the stellar light distribution, implying that it has originated from the mass loss of evolved stars.
- The diffuse gas surface brightness profile of NGC 5866 was well fitted by the standard beta model with the core radius equal to $9''.85$ and the slope parameter equal to $\beta \sim 0.62$.
- A substantial component of X-ray emission is found to originate from the discrete X-ray binary sources which are distributed throughout the galaxy. A total of 22 discrete sources have been detected within the optical $D_{25}$ extent of the galaxy.
- The higher values of the hardness ratios of the resolved sources imply that they are relatively harder and exhibit more complex emission characteristics compared to a single power law component. The combined spectrum of the X-ray emission from all the resolved sources was well constrained by an absorbed power law component with an additional thermal APEC component with the power law index parameter of 1.82.
- Deep broadband images in optical passbands reveal a prominent dust lane running along optical major axis of this galaxy. Morphology of dust closely matches with that of the ionized gas as well as hot X-ray emitting gas, pointing towards their common origin.
- The dust extinction curve derived for this galaxy delineates that physical properties of dust grains in the hostile environment of NGC 5866 are identical to the Milky Way dust grains; however, physical sizes of the grains are significantly modulated due to the sputtering and erosion by high energetic charged particles.

**Acknowledgement**:

The authors are thankful to IUCAA Girawali Observatory Team for providing technical help during observations and the IGT for telescope time on IGO 2-m telescope. This work has made use of *Chandra* archival X-ray data provided by the *Chandra X-ray Center*. The usage of databases like, Two Micron All Sky Survey (*2MASS*), NED, SDSS, ESO are gratefully acknowledged. SSS is very much thankful to University Grants Commission, New Delhi, India for providing financial support under the Minority Fellowship Program.